\documentclass[11pt, a4paper, twocolumn]{article}
\usepackage[dvips]{graphicx}
\usepackage[varg]{txfonts}

\usepackage{amssymb}
\usepackage{graphicx, color}
\usepackage{multirow}
\usepackage{color}

\newcommand{\OMIT}[1]{%
}

\newtheorem{Lemma}{Lemma}

\newtheorem{Assumption}{Assumption}

\newcommand{\Vol}{{\rm Vol}}
\newcommand{\shuffleN}{N_{{\rm shuffle}}}
\newcommand{\sampleN}{N_{{\rm sample}}}

\title{A Faster Method for Computing Gama-Nguyen-Regev's Extreme Pruning Coefficients}
\author{Yoshinori Aono
	\thanks{National Institute of Information and Communications Technology, Japan}
}

\begin{document}
\maketitle
\begin{abstract}
	This paper considers Gama-Nguyen-Regev's strategy \cite{GNR10} for optimizing pruning coefficients for lattice vector enumeration.
	We give a table of optimized coefficients and proposes a faster method for computing near-optimized coefficients for any parameters
	by interpolation.
\end{abstract}


\section{Introduction}

For given linearly independent vectors $\bold{b}_1,\ldots,\bold{b}_n$ in a Euclidean space,
the {\it lattice spanned by the vectors} is defined as the set of all integer coefficients:
$\mathcal{L}(\bold{b}_1,\ldots,\bold{b}_n)=\{ \sum_{i=1}^n a_i\bold{b}_i : a_i \in \mathbb{Z} \}$.
Its linear structure as a discrete subgroup of a Euclidean space
is used in many areas of science.
In particular, its computational properties have been used for investigating 
computational complexity and applications in cryptography.

In cryptography, it is interested in 
the hardness of several problems on lattices,
such as 
the shortest vector problem (SVP) whose goal is to find the non-zero lattice point nearest the origin,
and the closest vector problem (CVP) to find the nearest lattice point to a given target point.
It is important to estimate the practical hardness of these problems for cryptanalyses.
For this purpose, many algorithms have been proposed for solving SVP, CVP and several related problems.
Lattice reduction algorithms in particular are widely investigated.

The current strongest algorithm of this type is Chen-Nguyen's BKZ 2.0 algorithm \cite{CN11}
which improves on Schnorr's BKZ algorithm \cite{SE94,SH95}.
The basic strategy of BKZ-type algorithms is (i) pick a projected sublattice of a certain block size,
(ii) find short vectors in the sublattice, and (iii) update the entire lattice by the found vectors.
To find short vectors, Kannan's \cite{Ka83} {\it lattice enumeration algorithm} is usually used in step (ii).,
which is the heaviest part of BKZ algorithms.

Several efforts have been made to reduce the computational cost for lattice enumeration.
Since Schnorr-Euchner's pruning idea \cite{SE94} for bounding the projective lengths of vectors in the searching space,  
there has been considered to be a trade-off between cost and probability for finding a short vector.
Recently, Gama-Nguyen-Regev \cite{GNR10} proposed {\it the extreme pruning} technique 
that achieves dramatic improvement compared to Schnorr's pruning techniques.
However, 
since it requires a certain cost to optimize pruning coefficients besides lattice enumeration,
several previous works such as  \cite{KSD+11}, 
used the scaled version of the pruning coefficient of \cite{GNR10}, which is easily computable, in their experiments,
so these results are not sharp.
This paper gives a table of optimized coefficients for several parameters
and a method to give coefficients for any parameter by interpolation.
The expectation is that the coefficients of this paper can be used to so that results and consequent works will be improved.

\medskip

\noindent
{\bf Contribution}:
We follow the procedure for computing optimized pruning coefficients in lattice enumeration given by Gama-Nguyen-Regev \cite{GNR10},
and give a table of optimized coefficients for lattice basis satisfying Schnorr's geometric series assumption \cite{Sc03}
of several combinations of the following parameters:

$\bullet$ $n$, the lattice dimension,

$\bullet$ $\delta$, the $n$-th root of Hermite factor of lattice basis, and

$\bullet$ $p_0$, success probability

\noindent
are given.
Using optimized coefficients, 
we also propose a method for generating near-optimized coefficients for other parameters
by interpolation.
Differ from the original method,
our generating method does not require heavy Monte-Carlo procedure 
for generating coefficients.
In a reasonable dimension, it can be performed within a few seconds.

\medskip

\noindent
{\bf About this version}:
From the first version published in 2014, we inserted new Section~\ref{sec:newsection}
to  introduce our recent technique to compute approximations of 
enumeration cost (\ref{eqn:approxvol}) and success probability (\ref{eqn:approxprob});
both are completed in $O(n^2)$ floating point operations where $n$ is the lattice dimension.

For readers who are interested in this topic, 
we keep the descriptions of our heuristic optimization method in Section 4 
although they are outdated now.

\medskip

\noindent
{\bf Outline of this paper}:
Section~\ref{sec:preliminaries}, gives several definitions, notations, basic algorithms,
and heuristic assumptions.
Section~\ref{sec:approxes} introduces Gama-Nguyen-Regev's method
to approximate volume factors and success probabilities.
Section~\ref{sec:optimize} describes the method used here to optimize the pruning coefficients.
Experimental results are given in Section~\ref{sec:experiments}.
Finally, Section~\ref{sec:anyparams} proposes a new method 
to generate near-optimized pruning coefficients and comparison from scaled Gama-Nguyen-Regev coefficients.

\section{Preliminaries}

\label{sec:preliminaries}

For simplicity, sometimes $\bold{x},\bold{y},\ldots$ are used to denote vectors $(x_1,\ldots,x_n)$ in Euclidean spaces.
The dimension is from the context.
For a positive integer $n$, use $[n]$ to denote the set $\{ 1,\ldots, n\}$.
$\Delta_{n}$ and $\overline{\Delta_n}$ are the standard $n$-simplex $\{\bold{x} \in \mathbb{R}^n : 0\ge x_i\ {\rm for}\ \forall i, x_1+\cdots+x_n \le 1\}$
and its upper surface $\{ \bold{x}\in \Delta_n : x_1+\cdots+x_n = 1\} $, respectively.
Define $B_n(R)$ as the $n$-dimensional ball of radius $R$,
and let $V_n(R):=\pi^{n/2}R^n / \Gamma(n/2+1)$ be its volume.
Denote the surface of unit $n$-ball and its positive part by $S^{n-1}$ and
$S^{n-1}_+ := \{ \bold{x} \in S^{n-1} : x_i \ge 0\ {\rm for }\ \forall i\}$.

The sequence $R_1,\ldots,R_n\in \mathbb{R}$ is the pruning coefficient that satisfies 
(i) monotonically increasing and (ii) $0\le  R_i \le 1$ for all $i$.
For readability, define $\bold{r}=(R_1,\ldots,R_n)$.

\medskip

\noindent
{\bf Probability distributions}:~~
For a random variable $X$,
the notation $x \sim X$ means that $x$ is randomly sampled according to $X$.
For a set $Z$, $z\leftarrow Z$ means that $z$ is uniformly sampled from $Z$.
Gaussian distribution $N(\mu,\sigma)$ is the random variable with the probability density function (pdf) $\frac{1}{\sqrt{2\pi} \sigma}e^{-(x-\mu)^2 / 2\sigma^2}$ for all $x\in \mathbb{R}$.
The Dirichlet distribution $Dir(\alpha_1,\ldots,\alpha_{k})$ of order $k\ge 2$ 
has the pdf proportional to $\prod_{i=1}^k x_i^{\alpha_i-1}$ for $(x_1,\ldots,x_k) \in \overline{\Delta_k}$;
when $\alpha_i=1$ for all $i$, it is the uniform distribution over the simplex.

For $(X_1,\ldots,X_k) \sim Dir(\alpha_1,\ldots,\alpha_{k})$ and $(Y_1,Y_2) \sim Dir(\beta_1,\beta_2)$ where $\beta_1+\beta_2=1$,
the $k+1$ dimensional compounded random variable $(Y_1X_1,Y_2X_1,X_2,\ldots,X_k)$ has the distribution $Dir(\beta_1\alpha_1,\beta_2\alpha_1,\alpha_2,\ldots,\alpha_{k})$.

By the direct calculation of the probability density functions,
the following lemma holds.

\begin{Lemma}
 \label{lem:ball-convex} 
Suppose $(y_1,\ldots,y_n)$ distributes uniformly over $S^{n-1}$, 
then $(y_1^2,\ldots,y_n^2)$ follows $Dir(1/2,\ldots,1/2)$.
Conversely, 
	if $(z_1,\ldots,z_n) \sim Dir(1/2,\ldots,1/2)$,
	then its component-wise square rooted vector $(\sqrt{z_1},\ldots,\sqrt{z_n})$
	distributes uniformly over $S^{n-1}_+$.
\end{Lemma}

\medskip

\noindent
{\bf Cylinder-intersection-related objects}:~~
Fix a vector $\bold{r} = (R_1,\ldots,R_n) \in [0,1]^n$.
For an integer $k \in \{ 2,\ldots,n\}$, 
the {\it $k$-dimensional cylinder intersection} is defined by
\begin{equation}
	\label{eqn:k-baum}
	\begin{array}{l}
	\!\!\!	C_k(\bold{r}) = \left\{ \bold{x} \in \mathbb{R}^k :  \sum_{i=1}^\ell x^2_i \le R^2_\ell \ {\rm for}\ ^\forall \ell\in [k] \right\}.
	\end{array}
\end{equation}

\noindent
For even $k$, define
the {\it $k$-dimensional even-cylinder intersection} by
\begin{equation}
	\label{eqn:k-baum2}
	\begin{array}{l}
	\!\!\!\!\!\!\!\!\!\!\!\!\! C_k'(\bold{r}) = \left\{ \bold{x} \in \mathbb{R}^k :  \sum_{i=1}^\ell x^2_i \le R^2_\ell \ {\rm for}\ ^\forall \ {\rm even}\ \ell\in [k] \right\}.
	\end{array}
\end{equation}
Also, for $k'=k/2$, the corresponding convex bodies are defined by
\begin{equation}
	\label{eqn:k-simp}
	\begin{array}{l}
	\!\!\!\!\!\!\!\!\!\!\!
	\Delta_{k'}(\bold{r}) = \left\{ \bold{y} \in \Delta_{k'} 
		:  \sum_{i=1}^\ell y_i \le R^2_{2\ell} \ {\rm and}\ 
		{\rm for}\ ^\forall \ell \in [k'] \right\}
	\end{array}
\end{equation}
and
\begin{equation}
	\label{eqn:k-simp2}
	\begin{array}{l}
	\!\!\!\!\!\!\!\!\!\!\!\!\!
	\widetilde{\Delta_{k'}}(\bold{r}) = \left\{ \bold{y} \in \Delta_{k'} 
		:  \sum_{i=1}^\ell y_i \le R^2_{2\ell-1} \ {\rm and}\ 
		{\rm for}\ ^\forall \ell \in [k'] \right\}.
	\end{array}
\end{equation}
Note that it can be easily seen that $\Delta_{k'}(\bold{r})$ and  $\widetilde{\Delta_{k'}}(\bold{r})$ are convex sets
and $C_k(\bold{r})  \subset C_k'(\bold{r})$.


\medskip

\noindent
{\bf Spline interpolation of degree three}:~~
For two sequences of points $(x_0,\ldots,x_m)$ and $(y_0,\ldots,y_m)$ such that 
$x_0<\cdots<x_m$,
a procedure of spline interpolation returns a sequence of polynomials $f_j(x)$ defined in intervals $[x_{j-1},x_j]$ for $j=1,\ldots,m$.
Each polynomial satisfies (i) interpolating property: $f_j(x_{j-1})=y_{j-1}$ and $f_j(x_{j}) = y_j$, and
(ii) continuity of the first and second differential coefficients: $f'_{j}(x_j)=f'_{j+1}(x_j)$ and $f''_{j}(x_j)=f''_{j+1}(x_j)$.
For an efficient algorithm, see \cite{Sp95}.

\subsection{Hit-and-run Algorithm: Generating Random Points from Cylinder Intersections}

\label{sec:samplefromcylinder}

Smith \cite{Sm84}  proposed a random-walk-based method to generate random points from a convex body.
The algorithm, outline is given in Fig.~\ref{fig:hit-and-run}, is used to sample from simplexes (\ref{eqn:k-simp}) and (\ref{eqn:k-simp2}) to approximate the volume of cylinder intersections.

\begin{figure}[htbp]
\fbox{
\begin{minipage}{0.45\textwidth}
\begin{tabular}{lp{0.75\textwidth}}
{\bf Input} & Constraints for defining finite convex body $K$ in $\mathbb{R}^k$ \\
			& Initial point $\bold{y}_0 \in K$. \\
{\bf Output} & Randomly sampled point $\bold{y} \in K$ \\
Step 1:& Randomly choose a unit vector $\bold{d} \in S^{n-1}$ \\
Step 2:& Compute the minimum and maximum of $t\in \mathbb{R}$ so that $\bold{y}_0 + t\bold{d} \in K$, let them be $t_{{\rm min}}$ and $t_{{\rm max}}$  \\
Step 3:& Choose $t$ randomly from $[t_{{\rm min}},t_{{\rm max}}]$ \\
Step 4:& {\bf return} $\bold{y} = \bold{y}_0 +  t\bold{d}$ \\
\end{tabular}
\end{minipage}
}
\caption{Hit-and-run algorithm for generating random points}
\label{fig:hit-and-run}
\end{figure}

\medskip
\noindent
{\bf Implementing techniques}:
In step~1,
Marsaglia's method \cite{Ma72} is used to sample from the unit $k$-sphere:
Sample $d_i$ from the continuous Gaussian $N(0,1)$ independently for $i=1,\ldots,k$,
and then output $\bold{d} = (d_1,\ldots,d_k)/\sqrt{d_1^2+\cdots+d_k^2}$.

In Step~2,
since the target is a convex body defined by linear constraints,
it can use the improvement by Boneh-Caron \cite{BC90}
to speed up the sampling.
Concretely,
the maximum and minimum $t$ so that $\bold{y} + t\bold{d} \in K = \{ \bold{x} \in \mathbb{R}^k : \langle \bold{c}_i, \bold{x}\rangle \le w_i \}$ 
is given by
\[
	\max_{i: \langle \bold{c}_i,\bold{d} \rangle < 0} \left( \frac{w_i- \langle \bold{c}_i,\bold{y} \rangle }{\langle \bold{c}_i,\bold{d} \rangle} \right)
	\le t \le
	\min_{i: \langle \bold{c}_i,\bold{d} \rangle > 0} \left( \frac{w_i- \langle \bold{c}_i,\bold{y} \rangle }{\langle \bold{c}_i,\bold{d} \rangle} \right).
\]

\medskip
\noindent
{\bf Uniform sampling from even cylinder intersections}:
First the method to sample from the convex $\Delta_{k'}(\bold{r})$,
let it denote by {\tt ConvexSampler}$(k';\bold{r})$, is given.
The implementation here starts at a point $\bold{y}_0 = t\cdot (1,\ldots,1) \in K$ for some $t$,
and then execute the procedure in Fig.~\ref{fig:hit-and-run} $\shuffleN$ times to take a random initial point.
Next,
output $\sampleN$ points as a sequence of random points in $K$.
In Section~\ref{sec:approxes},
$\sampleN$ and $\shuffleN$ are experimentally chosen to sufficiently approximate the volume.

Next, construct {\tt EvenCylinderSampler}$(k;\bold{r})$ that 
samples uniformly from (\ref{eqn:k-baum2}).
Here, it is assumed $k$ is even and let $k'=k/2$.
Let $(y_1,\ldots,y_{k'})$ be a point generated by {\tt ConvexSampler}$(k';\bold{r})$.
Then, sample $\theta_i \leftarrow [0,2\pi]$ for $i\in [k']$ independently and uniformly and return
$(\sqrt{y_1}\cos\theta_1,\sqrt{y_1}\sin\theta_1,\ldots,\sqrt{y_{k'}}\cos\theta_{k'},\sqrt{y_{k'}}\sin\theta_{k'})$
$ \in \mathbb{R}^{k}$.

The correctness of this algorithm is considering the uniform distribution over $\Delta_{k'}$ and rejection sampling.
For $(y_1,\ldots,y_{k'}) \leftarrow \Delta_{k'}$,
the extended vector $(y_1,\ldots,y_{k'},1-y)$ where $y=y_1+\cdots+y_{k'}$ has the distribution $Dir(1,\ldots,1)$,
and the probability that $y_1+\cdots+y_{k'}=r$ is proportional to $r^{k'-1}$.
On the other hand, when $\theta$ distributes uniformly in the range $[0,2\pi]$,
$(\cos \theta,\sin \theta)$ is the uniform over $S^1$
and $(\cos^2 \theta,\sin^2 \theta) \sim Dir(1/2,1/2)$ by Lemma~\ref{lem:ball-convex}.
Thus, for uniformly and independently sampled $\theta_i \in [0,2\pi]$,
the compounded random variable
$(y_1\cos^2\theta_1,y_1\sin^2\theta_1,y_2\cos^2\theta_2,\ldots,y_{k'}\sin^2\theta_{k'},(1-y)\cos^2\theta_{k'+1},(1-y)\sin^2\theta_{k'+1}) \in \mathbb{R}^{k+2}$
has the distribution $Dir(1/2,\ldots,1/2)$.
Again by Lemma~\ref{lem:ball-convex},
its component-wise square rooted vector distributes uniformly over $S_+^{k+1}$.
Extracting the first $k$ coordinates $(\sqrt{y_1}\cos \theta_1,\sqrt{y_1}\sin\theta,\ldots,\sqrt{y_{k'}}\sin\theta)$,
the uniform distribution over the positive part of the unit ball is obtained
since its norm is $r=y_1+\ldots+y_{k'}$.
Considering rejection sampling,
the {\tt EvenCylinderSampler} outputs uniform points over $C'_k(\bold{r})$.

\subsection{Exact Computation of Volume of Truncated Simplexes}

\label{app:exactcomp}

Following Gama-Nguyen-Regev's original paper
and Chen-Nguyen \cite[Lemma A.1]{CN11},
we give the method to compute the volume of truncated simplex (\ref{eqn:k-simp}) exactly.
For readability, define the target object by 
\begin{equation}
	\label{eqn:def_tn}
	\begin{array}{l}
	\hspace{-6mm} T_{n} = \left\{ \bold{x}\in \Delta_n : C_\ell \le b_{\ell} \ {\rm for}\ \forall \ell\in [n] \right\}
	\end{array}
\end{equation}
for monotonically increasing series $b_1,\ldots,b_n \in [0,1]$,
and set $C_\ell := \sum_{i=1}^\ell x_i$.
The strategy is to express the volume by an integral of volume factors of cross sections in low dimensions,
and construct an inductive formula.

For $y\in [0,b_n]$,
denote  $T_{n-1}(y)$ the cross section of $T_n$ at $x_n=b_n-y$,
and let its volume by  $D_{n-1}(y)$.
Let $j$ be the index satisfying $y\in [b_j,b_{j+1})$; here it is assumed that $b_0=0$.
Since $C_n \le b_n$ implies $C_{n-1}\le y$,
some conditions in (\ref{eqn:def_tn}) are merged and $T_{n-1}(y)$ is explicitly written as
$\{ (x_1,\ldots,x_{n-1}) \in \mathbb{R}^{n-1}_+: C_1\le b_1,\ldots,C_j\le b_j, C_{n-1}\le y \}$.
With a similar argument, the cross section of $T_{n-1}(y)$ at $x_{n-1}=y-z$ for $z \le y$
is $\{ (x_1,\ldots,x_{n-2}) \in \mathbb{R}^{n-2}_+ : C_1\le b_1,\ldots,C_k\le b_k, C_{n-2}\le z \}$
for any $y \in [0,1]$ and $z\in [b_k,b_{k+1})$.
Since this set is not parametrized by $y$, it can be denoted as $T_{n-2}(z)$.
Therefore, for fixed $m<n$, $w$, $j$ such that $w\in [b_j,b_{j+1})$,
and any sequence $b_n \ge z_n \ge z_{n-1} \ge \cdots \ge z_{m+2} \ge w$,
the cross section of $T_n$ at hyperplanes $x_n=b_n-z_n$, $x_n+x_{n-1}=b_n-z_{n-1},\ldots,x_n+\cdots+x_{m+1}=b_n-w$,
are the same set
$T_{m}(w) := \{ (x_1,\ldots,x_{m}) \in \mathbb{R}^{m}_+ : C_1\le b_1,\ldots,C_j\le b_j, C_{m}\le \min(w,b_m) \}$.

Thus, letting $D_m(y)$ as the volume of $T_m(y)$,
it is easy to see
\[
	D_{i+1}(y) = \int^{y}_0 D_i(z) dz,
\]
which gives the explicit procedure to compute the target volume.
For $0\le j \le i \le n$,
let $F_{i,j}(y)$ be the function $D_i(y)$ in the range $y\in [b_j,b_{j+1})$.
For the base case $i=1$, since $T_1$ is a line,
the volume is $F_{1,0}=y$ for $y\in [0,b_1]$ and $F_{1,1}=b_1$ for $y\in [b_1,b_n]$.
For $j=0$ and $i>0$,
since $T_i(y)$ is a scaled standard simplex for $y\le b_1$,
$F_{i,0} = y^i/i!$ holds.
The general induction formula is given by
\begin{equation}
	\label{eqn:induction}
	\begin{array}{l}
		F_{i,j}(y) \\
		\displaystyle = \int^{b_1}_0 F_{i-1,0}(z)dz + \cdots + \int^{b_{j}}_{b_{j-1}} F_{i-1,j-1}(z)dz \\
		\displaystyle  + \int^{y}_{b_j} F_{i-1,j}(y) dz = F_{i,j-1}(b_{j})+ \int^{y}_{b_j} F_{i-1,j}(z)dz.
	\end{array}
\end{equation}
Using this,
the target volume is given by $\Vol T_n = F_{n,n}$.
Since it can be seen that $F_{i,j}(y)$ is a polynomial of degree $i-j$, 
total computation is done in $O(n^3)$ floating-point arithmetic operations.

With this technique, the probability 
\[
\Pr_{\bold{x} \leftarrow \Delta_{k'} } \left[ \bold{x}\in \Delta_{k'}(\bold{r}_{k}) \right] = k'! \cdot \Vol\Delta_{k'}(\bold{r}_{k})
\]
is also computable.

In preliminary computer experiments,
double precision computation has some errors for $n>75$.
We use the {\tt quad\_float} type in NTL library to compute the volume.

\subsection{Lattice Vector Enumeration Algorithm and Heuristic Assumptions}

\noindent
{\bf Notations}:~
For the lattice basis $\bold{b}_1,\ldots,\bold{b}_n$,
its Gram-Schmidt basis $\bold{b}^*_1,\ldots,\bold{b}^*_n$ and Gram-Schmidt coefficients $\mu_{i,j}$
are defined by $\bold{b}^*_1 := \bold{b}_1$ and for $i=2,\ldots,n$,
\[
	\bold{b}^*_i := \bold{b}_i - \sum_{j=1}^{i-1} \mu_{i,j} \bold{b}^*_j\
	{\rm and}\
	\mu_{i,j} := \frac{\langle \bold{b}_i,\bold{b}^*_j \rangle }{\langle \bold{b}^*_j,\bold{b}^*_j \rangle}.
\] 
The determinant, or the lattice volume, is defined by $\det(\mathcal{L}) := \prod_{i=1}^n ||\bold{b}^*_i||$.
For a vector $\bold{v}=\sum_{i=1}^n \alpha_i \bold{b}^*_i$, its $k$-th projection with respect to the basis is defined by 
$\pi_k(\bold{v}) = \sum_{i=k}^n \alpha_i \bold{b}^*_i$.

\medskip
\noindent
{\bf Enumeration algorithms}:~
The lattice enumeration algorithm for SVP and CVP for general dimension was first proposed by Kannan \cite{Ka83},
and modified by Schnorr-Euchner \cite{SE94} for proposing the BKZ lattice reduction algorithm.
The algorithm considers a search tree such that each node at depth $k$
is labeled by a lattice vector of the form $\sum_{n-k+1}^n a_i\bold{b}_i $ where $a_i\in \mathbb{Z}$.
For a node of $\bold{v}$ at depth $k$, 
it has children with labels $\bold{v}+a_{n-k} \bold{b}_{n-k}$ whose projective length $||\pi_{n-k}(\cdot)||$
is shorter than the threshold $c\cdot R_{k}$.	
Here, $c$ is the pruning radius set as $||\bold{b}_1||$ \cite{SE94,SH95,GNR10} or from the Gaussian heuristic \cite{CN11},
$R_k \in [0,1]$ is called  the pruning coefficients.
Efficient pseudo-codes for this algorithm are given in \cite{SE94,GNR10}.

\medskip
\noindent
{\bf Heuristic assumptions}:~
To analyze and improve these algorithms, 
several assumptions are introduced in previous papers.
Following them, these assumptions are assumed in the rest of this paper.

\begin{Assumption}
	\label{assump:gh}
	({\bf Gaussian heuristic assumption \cite{SH95}})
	Let $\mathcal{L}$ and $S$ be a full-rank lattice and a connected $n$-dimensional object respectively,
	i.e., there is no $(n-1)$-dimensional subspace that contains each one of them. 
	Then the number of lattice points in $S$ is approximated by $\Vol(S)/\det(L)$.
\end{Assumption}

\noindent
The length of the shortest vector is estimated with the radius of $n$-dimensional ball whose volume is equal to the determinant.
This is called {\it the Gaussian heuristic of lattice} and denoted as
$GH(\mathcal{L}) := V_n(1)^{-1/n} \det(\mathcal{L})^{1/n}$.

\begin{Assumption}
	\label{assump:gsa}
	({\bf Schnorr's geometric series assumption, GSA \cite{Sc03}})
	Let $\bold{b}_1,\ldots,\bold{b}_n$ be an output of a lattice reduction algorithm.
	Then there exists a constant that depends only on algorithms $r\in (0,1)$, 
	and it holds that $||\bold{b}^*_i||^2/||\bold{b}_1||^2 = r^{i-1}$ for $i\in [n]$.
\end{Assumption}

\noindent
There also exists an algorithm-dependent constant $\delta$, 
which claims the first vector of reduced basis satisfying $||\bold{b}_1|| = \delta^n \det(\mathcal{L})^{1/n}$.
$\delta^n$ is called {\it the Hermite factor} \cite{GN08b}.
They are connected by the relation $r=\delta^{-4n/(n-1)}$ if the GSA holds.

\begin{Assumption}
	\label{assump:gnr2}
	({\bf \cite[Heuristic 2]{GNR10}})
	Let $\bold{b}_1,\ldots,\bold{b}_n$ and $\bold{v}$ be an output of a lattice reduction algorithm and 
	one of the non-zero shortest vector in the lattice, respectively.
	Then, writing $\bold{v}/||\bold{v}|| = \sum_{i=1}^n w_i \bold{b}^*_i / ||\bold{b}^*_i||$,
	the vector $(w_1,\ldots,w_n)$ distributes uniformly over $S^{n-1}$, if the input of the reduction algorithm is random.
\end{Assumption}

\medskip
\noindent
{\bf Computational cost for enumeration}:~
Using the above assumptions \ref{assump:gh} and \ref{assump:gnr2},
Gama-Nguyen-Regev \cite{GNR10} estimated 
the total number $T$ of processed nodes during the enumeration algorithm 
and success probability $p$ for finding a shortest vector 
for input lattice basis $\bold{b}_1,\ldots,\bold{b}_n$.
They are respectively given by 
\begin{equation}
	\label{eqn:approxvol}
	T = \frac{1}{2} \sum_{k=1}^n \frac{ c^k \cdot \Vol C_k(\bold{r}) }{\prod_{i=n-k+1}^n ||\bold{b}^*_i||}
\end{equation}
and 
\begin{equation}
	\label{eqn:approxprob}
	p = \Pr_{\bold{u} \leftarrow  c\cdot S^{n}} \left[\sum_{i=1}^{\ell} u^2_i <  (c\cdot R_\ell)^2 \ {\rm for}\ ^\forall\ \ell\in [n] \right].
\end{equation}

The goal is to compute the pruning coefficients $\bold{r}=(R_1,\ldots,R_n)$ 
such that minimizes $T$ subject to $p\ge p_0$ for several parameters
when $c=GH(\mathcal{L})$.

\subsection{Computing Environments}

To perform preliminary experiments,
we used
a computing server with four AMD Opteron 6276 (2.3 GHz) cores, which can run 64 threads in parallel,
and 128 GB memory.
We use GMP library version 5.1.0 and NTL library version 6.0.0,
and {\tt g++} version 4.7.1 to compile the code.

The optimization procedure was conducted by V-nodes on the TSUBAME 2.5 supercomputer at the Tokyo Institute of Technology.
Each node has a virtual computing environment based on Intel Xeon 5670 (2.93 GHz), and can run 16 thread in parallel.
We use GMP library of version 5.1.2 and NTL library of version 6.0.0,
and {\tt g++} version 4.3.4 to compile the code.

In both experiments, we set the compiler option {\tt -Ofast -funroll-loops -march=native -ffast-math}.

\section{Approximating Computational Cost and Success Probability}

\label{sec:approxes}

\noindent
{\bf Preliminarily experiment for the hit-and-run algorithm}:~~
To approximate $\Vol(C_k(\bold{r}))$ by the Monte-Carlo approach, 
uniformly random points need to be sampled from $C'_k(\bold{r})$ by {\tt EvenCylinderSampler}.
To approximate the volume enough, we experimentally choose $\sampleN$.

For several dimension $n$, we set the pruning coefficients as $R^2_i=i/n$ for $i=1,\ldots,n$
and compute the actual value $A_n$ with high accuracy by the averaged value of 100 executions of 
the algorithm setting $\shuffleN=10^6$ and $\sampleN=5\cdot 10^{8}$.

\begin{figure}[ht]
	\begin{center}
	\includegraphics[scale=0.2]{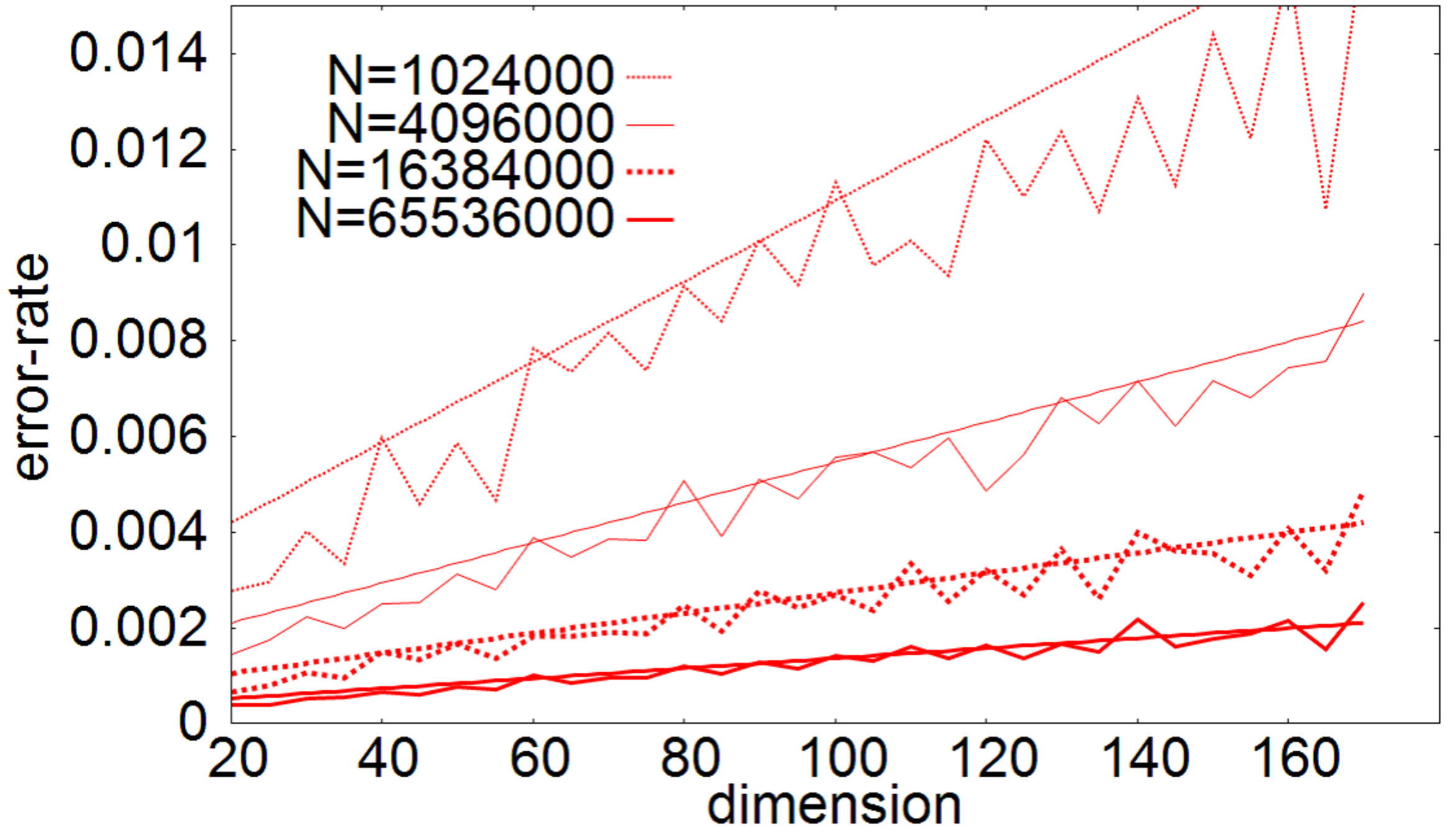}
	\caption{Experimental noise estimation of hit-and-run based volume estimation \label{fig:points_har}}
	\end{center}
\end{figure}

For these values, we measured the difference from approximated values with small $\sampleN$.
Fig.~\ref{fig:points_har} shows the average of error rate $|W_{n,N}-A_n|/A_n$ in 100 experiments,
where $W_{n,N}$ is an approximated value of dimension $n$ with enumerating $N=\sampleN$ points. 
We set $1.024\cdot 10^6$, $4.096\cdot 10^6$, $1.6384 \cdot 10^7$, and $6.5536\cdot 10^7$,
and $\shuffleN=10^5$ in this experiment.

The line charts and straight lines are respectively the averaged error and $0.085\cdot (n+30)/\sqrt{\sampleN}$
whose constants are selected so that errors are below with a reasonable probability.
Thus, to achieve the error-rate $\varepsilon$,
it needs to set
\begin{equation}
	\sampleN = 7.23\cdot 10^{-3} \times \left( \frac{n+30}{\varepsilon} \right)^2.
\end{equation}	
In the experiments below, $\sampleN=\max\{ 72.3 \times (dim+30)^2, 10^5 \}$ is used
so that  $\varepsilon=0.01$ if it does not mention the number of samples.

\subsection{Approximating Computational Cost for Lattice Enumeration}

\label{sec:approxcost}

To compute an approximated value of the total cost (\ref{eqn:approxvol}),
we introduce Gama-Nguyen-Regev's method and several implementing techniques.
Since $\Vol C_1(\bold{r})=2R_1$, interest here is in the situation where $k\ge 2$.

First, fix $k$ as an even number. 
Then $\Vol C'_k(\bold{r})$ is equal to
\[
	\hspace{-5mm}
	V_k(R_k)  \Pr_{\bold{x} \leftarrow B_{k}(R_k)} \left[ \bold{x}\in C'_k(\bold{r}) \right] 
	=	 V_k(R_k)   \Pr_{\bold{x} \leftarrow \Delta_{k'} } \left[ \bold{x}\in \Delta_{k'}(\bold{r}) \right]
\]
by the argument of uniform sampling in Section~\ref{sec:samplefromcylinder}.
The second probability is computable exactly in polynomial time by Section~\ref{app:exactcomp}.
Hence,  the desired volume is computed by the Monte-Carlo method with the formula
\begin{equation}
	\label{eqn:rel_l0}
\Vol C_k(\bold{r}) = \Vol C'_k(\bold{r}) \times \Pr_{x\leftarrow C'_k(\bold{r}) }\Big[x\in C_k(\bold{r}) \Big]
\end{equation}
with error $\varepsilon$.

Next consider the situation where the dimension $k$ is odd.
Suppose $\Vol C'_{k-1}(\bold{r})$ is computed.
Then we have  
\[
	\begin{array}{ll}
		\Vol C_k(\bold{r}) = &\Vol C'_{k-1}(\bold{r}) \cdot 2R_{k} \\
		&\times \Pr_{x\leftarrow C'_{k-1}(\bold{r}) \times [-R_{k},R_{k}] }\Big[x\in C_k(\bold{r}) \Big].
	\end{array}
\]
The error is $(1+\varepsilon)^2 -1 \approx 2\varepsilon$.
Here, to approximate the last probability,
the sampler takes the first $k-1$ coordinates from $C_{k-1}(\bold{r})$
and the last coordinate is from uniform distribution over the interval.

\medskip

\noindent
{\bf Implementing techniques}:
During the optimizing procedure,
interest here is in the situation where $T$ is smaller than current minimum cost $T_{{\rm cur}}$.
It can be observed that factors in the sigma 
\[
	\frac{c^k \Vol C_k(\bold{r}) }{\prod_{i=n-k+1}^n ||\bold{b}^*_i||}
\]
are a roughly concave function of $k$,
and the values for $0.2n\le k\le 0.8n$ are dominated.
For even $k$, $\Vol C_k(\bold{r})$ is bounded lower from $L_k := V_k(R_k)\cdot \Pr_{\bold{x} \leftarrow \Delta_{k'}} \left[ \bold{x}\in \widetilde{\Delta_{k'}}(\bold{r}) \right]$.
For odd $k\ge 3$, assuming the factors form a concave function of $k$, $\Vol C_k(\bold{r})$ is bounded lower by
$( L_{k-1} + L_{k+1})/2$.
Thus, it can be estimated a lower bound of the sum (\ref{eqn:approxvol}) and the process is aborted if it is larger than $T_{{\rm cur}}$
without performing the Monte-Carlo simulation.

Since computing all volume factors requires quite large costs,
a partial sum is considered to give a sufficiently accurate approximation.
That is, the computation can be aborted if the partial sum $\sum_{i=1}^{k'-1}$ exceeds $T_{{\rm cur}}$,
or the factor is sufficiently smaller the than partial sum $\sum_{i=1}^{k'-1}$.
in these experiments we set the condition that factor $<10^{-5}\cdot \sum_{i=1}^{k'-1}$.
In the latter case, sufficiently approximated value of the computational cost is computed.

\subsection{Probability Computation}

\label{sec:mainprobcomp}

We give the method
to approximate the probability of finding a shortest vector (\ref{eqn:approxprob}).
In Chen-Nguyen \cite{CN11}, they explained the fast computation of lower and upper bounds 
only in even dimensional case.
We extend the method for the situation where the dimension is odd.

First consider the situation where $n$ is even and $R_{2i}=R_{2i-1}$ for all $i\in [n/2]$,
and assume that $R_{n-1}=R_n=1$.
Then by the argument in Section~\ref{sec:samplefromcylinder},
\begin{equation}
	\label{eqn:upperprobability}
	\begin{array}{l}
	\Pr_{\bold{u} \leftarrow c \cdot S^{n-1}} \left[\sum_{i=1}^{\ell} u^2_i < (c \cdot R_\ell)^2 \ {\rm for}\ ^\forall\ \ell\in [n] \right] \\
	=
	\Pr_{\bold{x} \leftarrow Dir(1,\ldots,1)}  \left[  \sum_{i=1}^{\ell} x_i < R^2_{2\ell} \ {\rm for}\ ^\forall\ \ell\in [n/2] \right] \\
	=
	\Vol\Delta_{n/2-1}(\bold{r}).
	\end{array}
\end{equation}
The last probability is easily computed.
It is easy to see that the above is equal to the hyperarea $U := \Vol(S^{n-1} \cap C'_n(\bold{r}))$,
which gives an upper bound for 	(\ref{eqn:approxprob}).
Thus, for general $\bold{r}$ 
the desired probability is computed by the relation
\begin{equation}
	(\ref{eqn:approxprob}) = U \cdot \Pr_{\bold{x} \leftarrow C'_{n}(\bold{r})}
	\left[ \bold{x}\in C_{n}(\bold{r}) \right].
\end{equation}
The rigid lower bound is also easily computable by 	$L := \Vol \widetilde{\Delta}_{n/2-1}(\bold{r})$.

Next we consider the situation where $n$ is odd.
If $(u_1,\ldots,u_{n+1}) \sim S^n$, the vector 
\[
	\left( \frac{u_1}{\sqrt{1-u^2_{n+1}}}, \frac{u_2}{\sqrt{1-u^2_{n+1}}}, \ldots ,\frac{u_{n}}{\sqrt{1-u^2_{n+1}}} \right) \in \mathbb{R}^n
\]
uniformly distributes over $S^{n-1}$.
Hence,
\[
	\begin{array}{ll}
	(\ref{eqn:approxprob}) 
	&= \Pr_{\bold{u}\leftarrow S^n} \left[ \frac{\sum_{i=1}^\ell u^2_i}{1-u^2_{n+1}} \le R^2_\ell\ {\rm for}\ \forall \ell\in [n] \right]  \\
	&\le \Pr_{\bold{u}\leftarrow S^n} \left[ \sum_{i=1}^\ell u^2_i \le R^2_\ell\ {\rm for}\ \forall \ell\in [n] \right] 	 \\
	&\le \Pr_{\bold{u}\leftarrow S^n} \left[ \sum_{i=1}^{\ell'} u^2_i \le R^2_{2\ell'}\ {\rm for}\ \ell'\in [(n-1)/2] \right].
	\end{array}
\]
The last probability, again denoted as $U$, is easily computed and sampling from $C'_{n+1}(\bold{r})$ is also easy.
Thus, letting $R_{n+1}=1$, the desired probability is given by 
\begin{equation}
		(\ref{eqn:approxprob}) = U \cdot \Pr_{\bold{x} \leftarrow C'_{n+1}(\bold{r})}
     	 \left[ \frac{\sum_{i=1}^\ell u^2_i}{1-u^2_{n+1}} \le R^2_\ell\ {\rm for}\ \forall \ell\in [n] \right].
\end{equation}
This $U$ is also used as the rigid upper bound of the probability.

To give a rigid lower bound $L$,
consider a similar relation that 
if $(u_1,\ldots,u_{n-1}) \sim S^{n-2}$, and $v$ follows the probability distribution $V$ over $[-1,1]$ 
such that $\Pr[V=v]$ is proportional to $\left( \sqrt{1-v^2}\right)^{n-2}$,
the vector 
\[
	\left( \sqrt{1-v^2}u_1,\ldots,\sqrt{1-v^2}u_{n-1},v  \right)
\]
is also uniform distribution over $S^{n-1}$.
Hence, the following relation gives the lower bound:
\[
	\hspace{-5mm}
	\begin{array}{ll}
	(\ref{eqn:approxprob}) 
	& \displaystyle = \Pr_{\bold{u}\leftarrow S^{n-2}, v\leftarrow V} \left[ (1-v^2) \sum_{i=1}^\ell u^2_i \le R^2_\ell\ {\rm for}\ \forall \ell\in [n] \right]  \\
	& \displaystyle \ge \Pr_{\bold{u}\leftarrow S^{n-2}} \left[ \sum_{i=1}^\ell u^2_i \le R^2_\ell\ {\rm for}\ \forall \ell\in [n] \right]  \\
	& \displaystyle  \ge  \Pr_{\bold{u}\leftarrow S^{n-2} }  \left[ \sum_{i=1}^{\ell'} u^2_i \le R^2_{2\ell'-1}\ {\rm for}\ \forall \ell'\in [(n+1)/2] \right].
	\end{array}
\]

\medskip

\noindent
{\bf Implementing techniques}:~~
Since the focus here is situation where the probability is larger than the given value $p_0$ during the optimizing process, it can abort the computation if $p_0$ is not between $U$ and $L$.
Also the Monte-Carlo computation is also aborted
after sampling $\sampleN/100$ points
if the intermediate result (i.e., a rough approximation of the probability)
is not within the range  $[0.9,1.1]p_0$.

\subsection{How to Compute the Volumes Efficiently}

\label{sec:newsection}

We introduce our method to find the enumeration cost (\ref{eqn:approxvol}) and success probability (\ref{eqn:approxprob})
in $O(n^2)$ floating point operations.
They have implemented in the progressive BKZ library \cite{pbkzlib} newer than version 201803.

Recall that the inputs are $(R_1,\ldots,R_n)$ and $(\| {\bf b}^*_1 \| ,\ldots, \| {\bf b}^*_n\|)$
and suppose one wants to compute the upper bound of probability (\ref{eqn:upperprobability})
or the upper bound of the cost given by
\[
	\sum_{k=1}^{n/2} \frac{c^k \Vol(\Delta_k ({\bf r})) }{\prod_{i=n-2k+1}^n \| {\bf b}^*_i  \| }.
\]
Thus, it suffices to compute $\Vol (\Delta_1({\bf r})),\ldots, \Vol (\Delta_{n/2}({\bf r}) )$
which are equivalent to $F_{j,j}$ for $j=1,\ldots,n/2$ defined by (\ref{eqn:induction}).

Taking derivatives of both sides of (\ref{eqn:induction}) , we have $F'_{i,j}(y) = F_{i-1,j}(y)$ 
and
\begin{equation}
\label{eqn:Hderivative}
F^{(k)}_{i,j}(y) = F_{i-k,j}(y).
\end{equation}
Moreover, 
by substituting $x=b_j$ to (\ref{eqn:induction}), we have $F_{i,j}(b_j) = F_{i,j-1}(b_j)$
for any $i$ and $j$.

Considering the Taylor expansion of $F_{i,j}(y)$ at $y=b_j$,
we have 
\[
	\begin{array}{ll}
		F_{i,j}(y) & \displaystyle = \sum_{k=0}^{i-j} F_{i,j}^{(k)}(b_j) \cdot \frac{(y-b_j)^k}{k!} \\
		& \displaystyle =  \sum_{k=0}^{i-j} F_{i-k,j} (b_j) \cdot \frac{(y-b_j)^k}{k!} \\
		& \displaystyle =  \sum_{k=0}^{i-j} F_{i-k,j-1} (b_j) \cdot \frac{(y-b_j)^k}{k!}.
	\end{array}
\]

On the other hand,
considering the Taylor expansion of $F_{i,j-1}(x)$, we have 
\[
\begin{array}{ll}
	F_{i,j-1}(y) & \displaystyle = \sum_{k=0}^{i-j+1} F_{i-k,j-1}(b_j) \cdot \frac{(y-b_j)^k}{k!} \\
	& \displaystyle = F_{i,j}(y) + F_{j-1,j-1} \cdot \frac{(y-b_j)^{i-j+1}}{(i-j+1)!} 
\end{array}
\]

Again using (\ref{eqn:Hderivative}) we can easily see that $F_{j-1,j-1} / (i-j+1)!$ is the 
leading coefficient of $F_{i,j-1}$, let it be $h_{i,j-1}$.
Therefore, we obtain the induction formula 
\[
	F_{i,j}(y) = F_{i,j-1}(y) - h_{i,j-1} (x-b_j)^{i-j+1}. 
\]

Starting from $F_{d,0}=x^d/d!$, we can compute all $F_{d,0},\ldots,F_{d,d}$ 
in $O(d^2)$ floating number operations.

\section{Optimizing Pruning Coefficients}

\label{sec:optimize}

We give a method to compute optimized pruning coefficients
for several parameters $(n,\delta,p_0)$;
recall they are lattice dimension, $n$-th root of Hermite factor, and lower bound of success probability, respectively.
The task is to minimize the cost (\ref{eqn:approxvol}) subject to the probability (\ref{eqn:approxprob}) larger than $p_0$.

\medskip

\noindent
{\bf Outline of the procedure}:
To represent pruning coefficients, 
we use {\it defining points} $s_0,\ldots,s_m \in [0,1]$
which are extracted to the function $f(x)$ in the interval $[0,n]$ that passes through 
$(nm/i,s_i)$ for $i=0,\ldots,m$ by the spline interpolation. 
We set the pruning coefficient $R^2_j$ as the maximum value of $f(x)$ in the interval $[0,j/n]$,
and reset $R_j$ to 0 or 1 if $R_j<0$ or $R_j>1$, respectively.
In our experiment we set $m=16$.

By Schnorr's GSA, 
it is also assumed $||\bold{b}^*_i||=r^{(i-1)/2}$ where $r=\delta^{-4n/(n-1)}$.
The pruning radius is set as the Gaussian heuristic $c=GH(\mathcal{L})=V_n(1)^{-1/n}r^{(n-1)/4}$.

Starting from the linear function $s_i = i/m$, 
the procedure in Fig.~\ref{fig:outlines} searches the optimized function.
The subroutines ${\tt cost}(s_1,\ldots,s_m)$ and ${\tt prob}(s_1,\ldots,s_m)$ compute
the approximated computational cost and success probability of extracted coefficients, respectively.
${\tt perturbate}(s_1,\ldots,s_m)$ and ${\tt modifycurve}(s_1,\ldots,s_m)$ 
are functions to operate points described below.
In step 2, starting at $\eta=0.01$ and use the binary search
to find the smallest $\eta$ so that the probability is larger than $p_0$.
Note that the algorithm did not output the last $(s_0,\ldots,s_m)$ because 
$T_{{\rm cur}}$ is a local minimum cost updating by each execution of {\tt modifycurve}.
The procedure actually save the pairs of defining points and computing cost,
and finally outputs the best points.

After finishing the optimizing procedure, 
shift the curve by $s_i \leftarrow s_i + \eta$ for all $i$
so that the probability is very close to $p_0$.
In this postprocessing, we set $\sampleN=7230\times (n+30)^2$ so that $\varepsilon=0.001$.

\begin{figure}[h]
\fbox{
\begin{minipage}{0.45\textwidth}
\begin{tabular}{lp{0.75\textwidth}}
	{\bf Input} & $(n,\delta,p_0)$ \\
	{\bf Output} & $(s_0,\ldots,s_m)$ to define the optimized function  \\
	Step 1:  & Initialize $s_i = i/m$ for $i\in [m]$.  \\
	Step 2:  & {\bf if}  ${\tt prob}(s_1,\ldots,s_m)<p_0$ {\bf then} set  $s_i \leftarrow s_i + \eta$ for all $i$ so that ${\tt prob}(s_1,\ldots,s_n)> p_0$ \\
	Step 3: & Set $T_{{\rm cur}} \leftarrow {\tt cost}(s_1,\ldots,s_m)$ \\
	Step 4: & {\tt perturbate} $(s_1,\ldots,s_m)$ \\
	Step 5: & {\bf if } ${\tt cost}(s_1,\ldots,s_m) < T_{{\rm cur}}$ {\bf then} save the points  \\
			  & \ \ \ \ Update $T_{{\rm cur}} \leftarrow {\tt cost}(s_1,\ldots,s_m) $ \\
	Step 6: & {\bf if} $T_{{\rm cur}}$ is not updated in the current 20 perturbations \\
			& \ \ \ \  {\bf then} load the points achieving cost $T_{{\rm cur}}$\\
	Step 7: & {\bf if} $T_{{\rm cur}}$ is not updated in the current 50 perturbations \\
				& \ \ \ \  {\bf then} {\tt modifycurve} $(s_1,\ldots,s_m)$;  \\
				& \ \ \ \ \ \ \ \ $T_{{\rm cur}} \leftarrow {\tt cost}(s_1,\ldots,s_m)$; {\bf goto} Step 5\\ 
	Step 8: & {\bf if} $T_{{\rm cur}}$ is not updated in the current 200 perturbations \\
				& and there is no need to modify the curve {\bf then} finish the computation and output the saved points such that 
				 they give the minimum cost {\bf else goto} Step 4\\ 
\end{tabular}
\end{minipage}
}
\caption{Outline of optimization procedure\label{fig:outlines}}
\end{figure}

\medskip

\noindent
{\bf Perturbating subroutine}:
The procedure is given in Fig.~\ref{fig:optpertubate}. 
We assume that the probability of input points is larger than and sufficiently close to $p_0$.
In Step 1, the perturbation strategy $f=1$ means that increase $s_k$ to add the probability,
then decrease $s_{k+d}$ to reduce the probability;
strategy $f=-1$ is that of inverting increase and decrease.
The procedure $rand(\alpha,\beta)$ generates uniform random real number between $\alpha$ and $\beta$.

\begin{figure}[h]
\fbox{
\begin{minipage}{0.45\textwidth}
\begin{tabular}{lp{0.75\textwidth}}
{\bf Input} & ($s_0,\ldots,s_m$) such that ${\tt Prob}(s_1,\ldots,s_m)>p_0$ \\
			& Perturbation steps $\varepsilon$, $\varepsilon'$ \\
{\bf Output} & New defining points having probability larger than $p_0$\\
	Step 1: & $k \leftarrow \{ 0,\ldots,m \}$ // starting index\\
			& $d \leftarrow \{ -1,+1 \}$ //direction of index \\
			& $f \leftarrow \{ -1,+1 \}$ // perturbation strategy\\
Step 2:	& $s_k\leftarrow s_k + f \cdot rand(0,\varepsilon)$ \\
			& {\bf while} ($0\le k+d\le m$) {\bf do} \\
			& \ \ \ $k\leftarrow k+d$ \\
			& \ \ \ $p_{{\rm prev}} \leftarrow {\tt Prob}(s_1,\ldots,s_m)$ \\
			& \ \ \ $s_k \leftarrow s_k - f \cdot rand(0,\varepsilon')$ \\
			& \ \ \ $p_{{\rm curr}} \leftarrow {\tt Prob}(s_1,\ldots,s_m)$ \\
			& \ \ \ {\bf if}\ $(p_{{\rm prev}},p_{{\rm curr}})$ crosses $p_0$ {\bf then} \\
			& \ \ \ \ \ \ {\bf if} $f=-1$ {\bf then return} current $s_0,\ldots,s_m$ \\
			& \ \ \ \ \ \ \ \ \ {\bf else return} previous $s_0,\ldots,s_m$ \\
			& {\bf end-while} \\
Step 3:	& {\bf if} ${\tt Prob}(s_1,\ldots,s_m)>p_0$ {\bf then return} $s_0,\ldots,s_m$ \\
			& \ \ \ {\bf else} reset the points and {\bf goto} Step 1
\end{tabular}
\end{minipage}
}
\caption{Perturbation procedure\label{fig:optpertubate}}
\end{figure}

\medskip

\begin{figure}[h]
\includegraphics[scale=0.25]{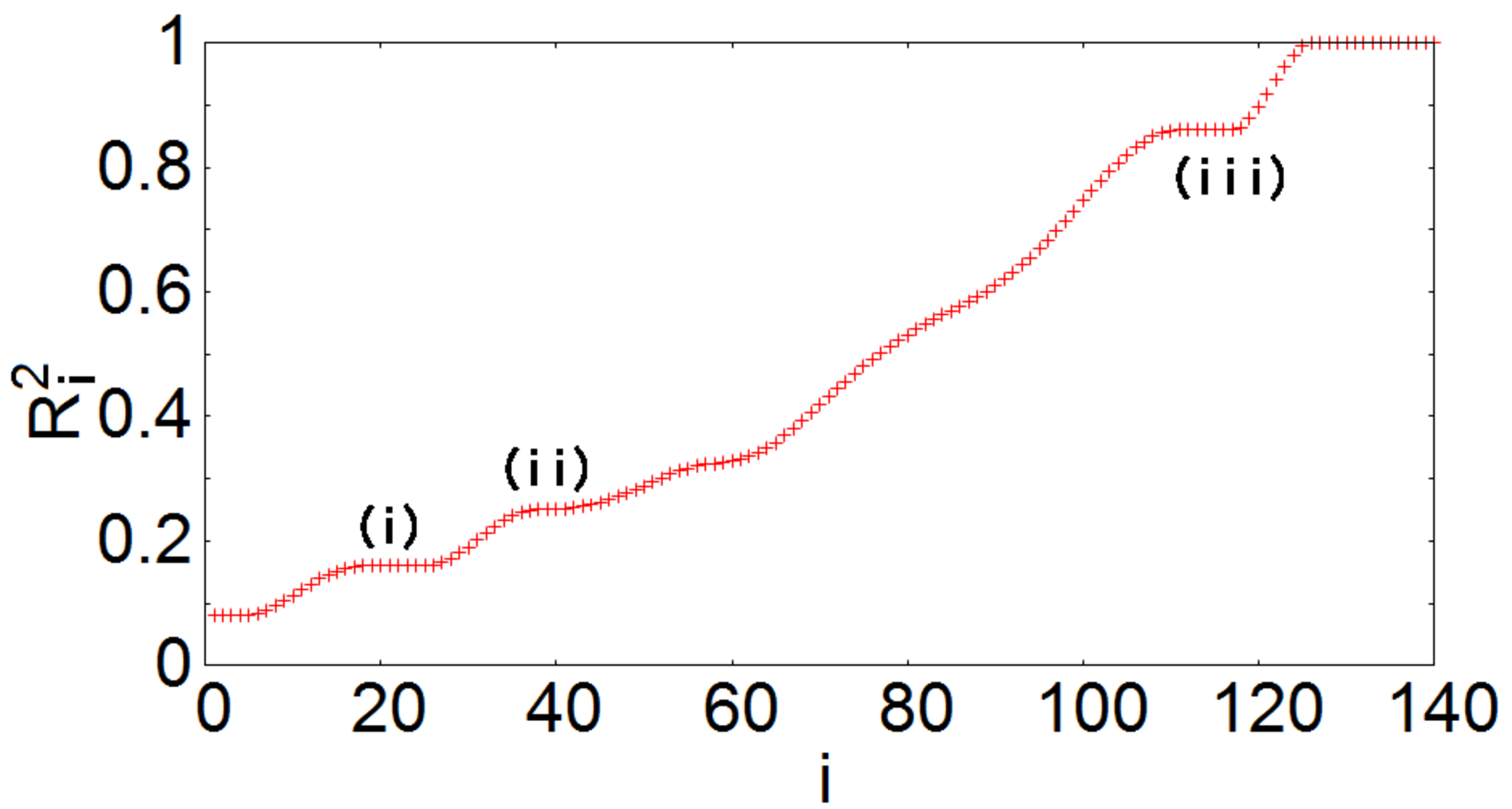}
\caption{ Example of nonsmooth curve \label{fig:strange_curve}}
\end{figure}

\begin{figure}[h]
\includegraphics[scale=0.25]{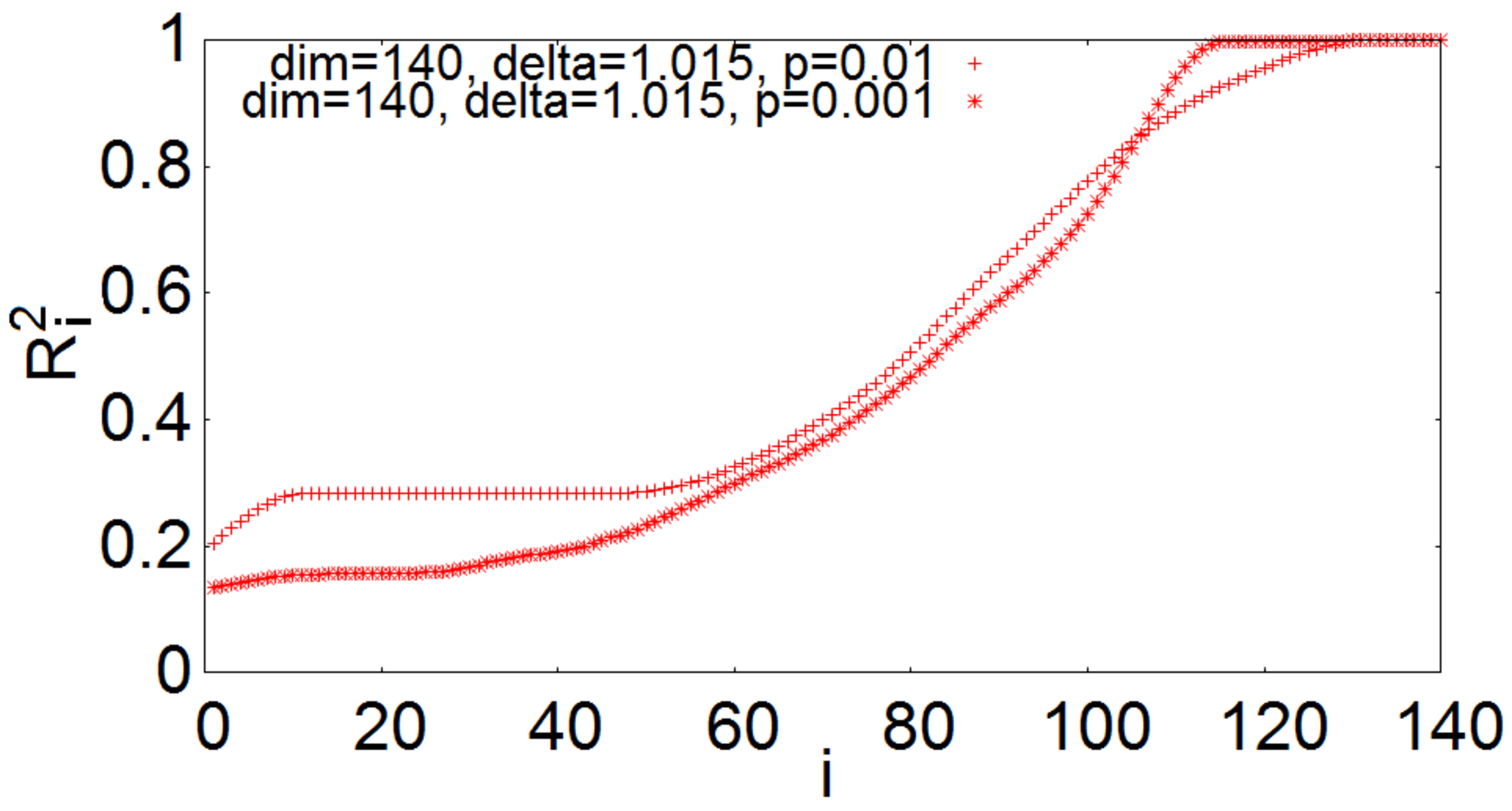}
\caption{ Example of styled curves \label{fig:styled_curve}}
\end{figure}

\noindent
{\bf Modifying subroutine}:
In our preliminary experiments for optimization, 
the shape of curve sometimes stays at a nonsmooth curve, such as in Fig.~\ref{fig:strange_curve}.
From the example curve in \cite{GNR10} and our final results in Fig.~\ref{fig:styled_curve},
we say that the curve is {\it smooth} if
at most one flat part exists 
and the (discrete) second derivative is positive in the medium part.
As shown in the figure, the nonsmooth curve has two or more horizontal parts ((i), (ii), and (iii)) under $R_i=1$
and several inflection points.

The procedure to modify the curve is given as follows.
For the recovered sequence $R_1^2,\ldots,R_n^2$, it starts from $i=1$ and increases $i$.
If there is a second flat part at $i$, i.e., there exists $i'<i$ such that $R_{i'-1}=R_{i'} < R_{i}=R_{i+1}$, except for $R_i=1$,
it modifies the defining points as $s_j \leftarrow s_j - 0.001$, where $j= \lfloor i\cdot m/N \rfloor$,
until the problem is solved.
Next, it simultaneously checks the discrete second derivative 
$R_{i+3}^2-R_{i}^2+R^2_{i-3}$.
If it is smaller than -0.005 within the band $R^2_i \in [0.2,0.9]$, 
modify the defining points as $s_j \leftarrow s_j - 0.001$ and  $s_{j+1} \leftarrow s_{j+1} + 0.001$.
In these modifications, it restricts the defining points so that $s_0,\ldots,s_m$ is monotonically increasing,
i.e., after changing $s_j$, do operations $s_j \leftarrow \max \{ s_{j-1},s_j\}$ and $s_j \leftarrow \min \{ s_{j+1},s_j\}$.
They are also restricted within the range $[0,1]$.

\section{Experimental Results}

\label{sec:experiments}

\begin{figure}[h]
\includegraphics[scale=0.25]{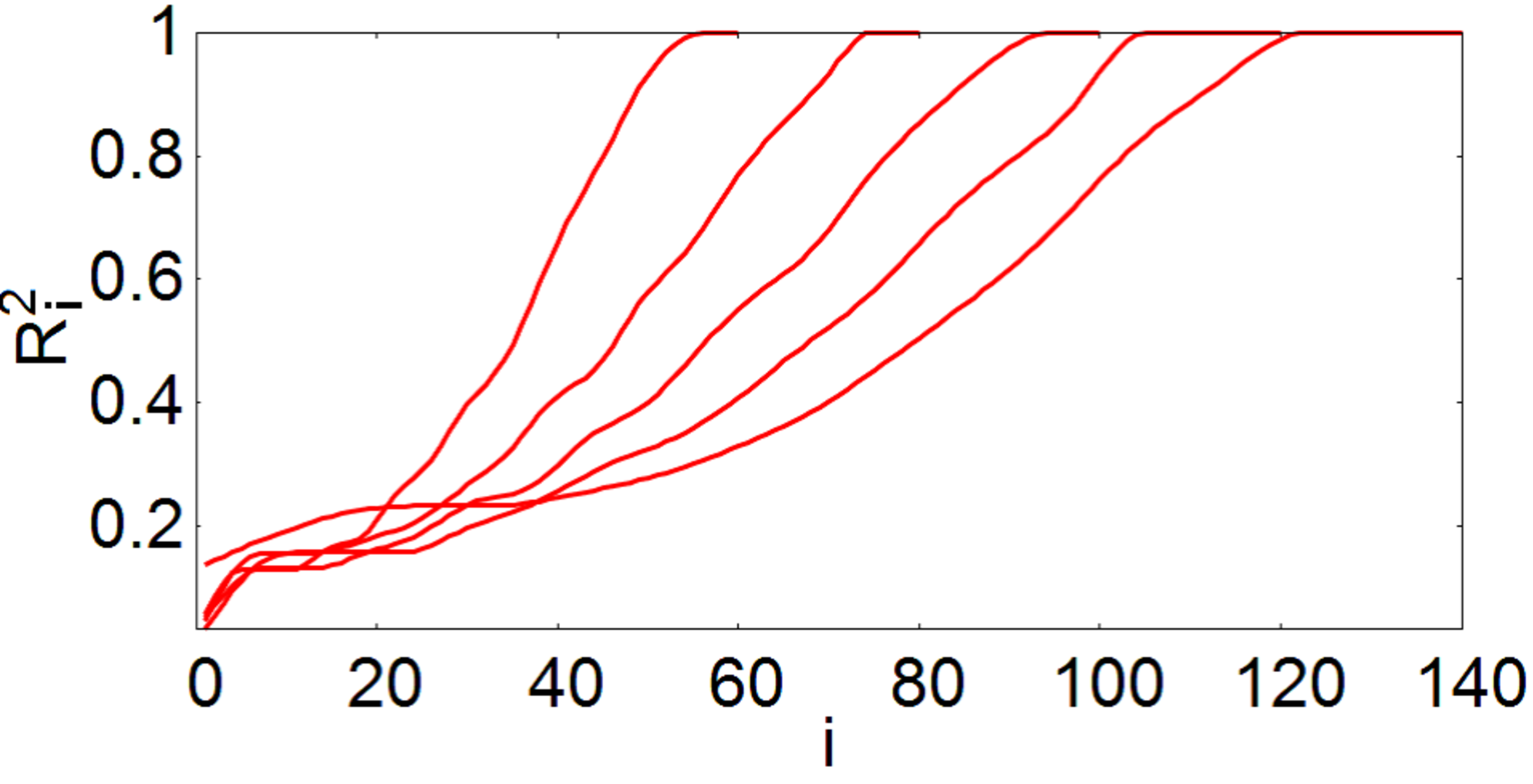}
\caption{Example of optimized curves for $\delta=1.01$, $p_0=0.01$ and $n=60,80,\ldots,140$ \label{fig:examplegraph}}
\end{figure}

Using the above strategy, 
we compute optimized pruning functions 
for parameters  
$\delta=1.02$, $1.015$, $1.01$, $1.005$, $n=60,80,\ldots,200$, and $p_0=10^{-e}$
for $e=1,2,3,4,6,12,24$.
As an example, Table~\ref{tab:examplesi} and Fig.~\ref{fig:examplegraph}
give the table of $s_i$ and graphs for $\delta=1.01$, $p_0=0.01$ and $n=60,80,\ldots,140$ respectively.
All values are published at \verb+http://www2.nict.go.jp/nsri/fund/aonodata/+\\
\verb+coefftable_ver01.pdf+.

\begin{table}
	\caption{Optimized defining points for $\delta=1.01$, $p_0=0.01$ and $n=60,80,\ldots,140$ \label{tab:examplesi}}
	\begin{center}
	\begin{footnotesize}
	\begin{tabular}{|l||l|l|l|l|l|}
		\hline
		& 60 & 80 & 100 & 120 & 140\\
		\hline 		\hline
		0 &0.0214&0.01641&0.0324&0.0098&0.1318\\
		\hline
		1&0.1208&0.1385&0.1270 &0.1437&0.1859\\
		\hline
		2&0.1208&0.1537&0.1287&0.1484&0.2240\\
		\hline
		3&0.1282&0.1604&0.1551&0.1484&0.2326\\
		\hline
		4&0.1642&0.1826&0.1878&0.1944&0.2336\\
		\hline
		5&0.1845&0.2099&0.2405&0.2354&0.2565\\
		\hline
		6&0.2510&0.2626&0.2648&0.2939&0.2871\\
		\hline
		7&0.3043&0.3201&0.3430&0.3380&0.3353\\
		\hline
		8&0.3884&0.4043&0.3950&0.4022&0.3978\\
		\hline
		9&0.4501&0.4606&0.4884&0.4897&0.4860\\
		\hline
		10&0.5550&0.5691&0.5753&0.5742&0.5808\\		
		\hline
		11&0.6788&0.6479&0.6480&0.6870&0.6936\\
		\hline
		12&0.7794&0.7548&0.7640&0.7829&0.8241\\	
		\hline
		13&0.8837&0.8432&0.8601&0.8759&0.9191\\	
		\hline
		14&0.9627&0.9200&0.9396&1.0000&1.0000\\	
		\hline
		15&1.0000&1.0005&1.0000&1.0000&1.0000\\	
		\hline
		16&1.0000&1.0007&1.0000&1.0000&1.0000\\	
		\hline
	\end{tabular}
	\end{footnotesize}
	\end{center}
\end{table}

\subsection{Discussion for Optimality}

The optimizing problem is known as a monotonic programming,
which is defined by the problem for searching $\arg\min\{ f(x) : g(x)\ge 0\}$
for monotonically increasing functions $f,g:\ \mathbb{R}^n \rightarrow \mathbb{R}$,
i.e.,
in component-wise meaning $x\le x'$ implies that $f(x)\le f(x')$ and $g(x)\le g(x')$.
Since our optimizing strategy is based on a random walk,
it does not guarantee the global optimality of the result.
Nevertheless, by the following observations, 
we tried to justify the result.

We performed preliminary experiments using $m=12$ defining points
for the parameter $(n,\delta,p_0)=(60,1.02,0.01)$.
Starting at $s_i=i/12$ (linear function) and $s_i=\sqrt{i/12}$ (curved function),
both are converged to a similar curve.

\begin{figure}
\includegraphics[scale=0.35]{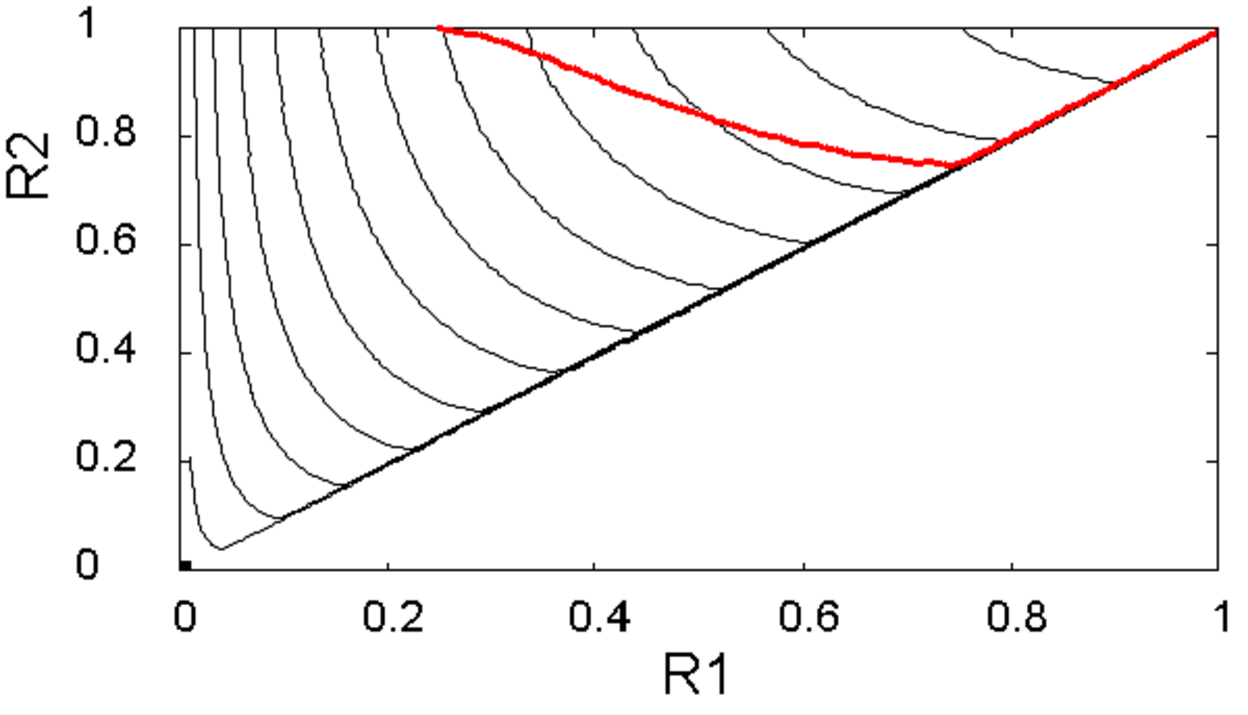}
\caption{ Contour lines of computational cost and success probability \label{fig:threedimexample}}
\end{figure}

For the dimension $n=3$,
Fig.~\ref{fig:threedimexample} shows 
the contour line of computational cost $T$ and success probability $p$
with setting $||\bold{b}^*_i|| = 0.8^{i-1}$ for $i=1,2,3$.
The thin lines are the contour of $T$ within 0.5 steps and the upper right side is about 6.5.
The bold line is contour of $p=0.5$.
It is known that the optimal point lies on the surface $\Pr=p_0$ \cite{Tu00} in monotonic programming.
Thus, the random walk procedure is expected to search points near the contour of $p=p_0$,
which does not have a local optimal on the bold line in the figure.

Therefore, we might assume that the algorithm always goes to the global optimal 
irrespective of starting curve.

\section{Generating Pruning Functions for any Parameters}

\label{sec:anyparams}

\begin{figure}[h]
\includegraphics[scale=0.25]{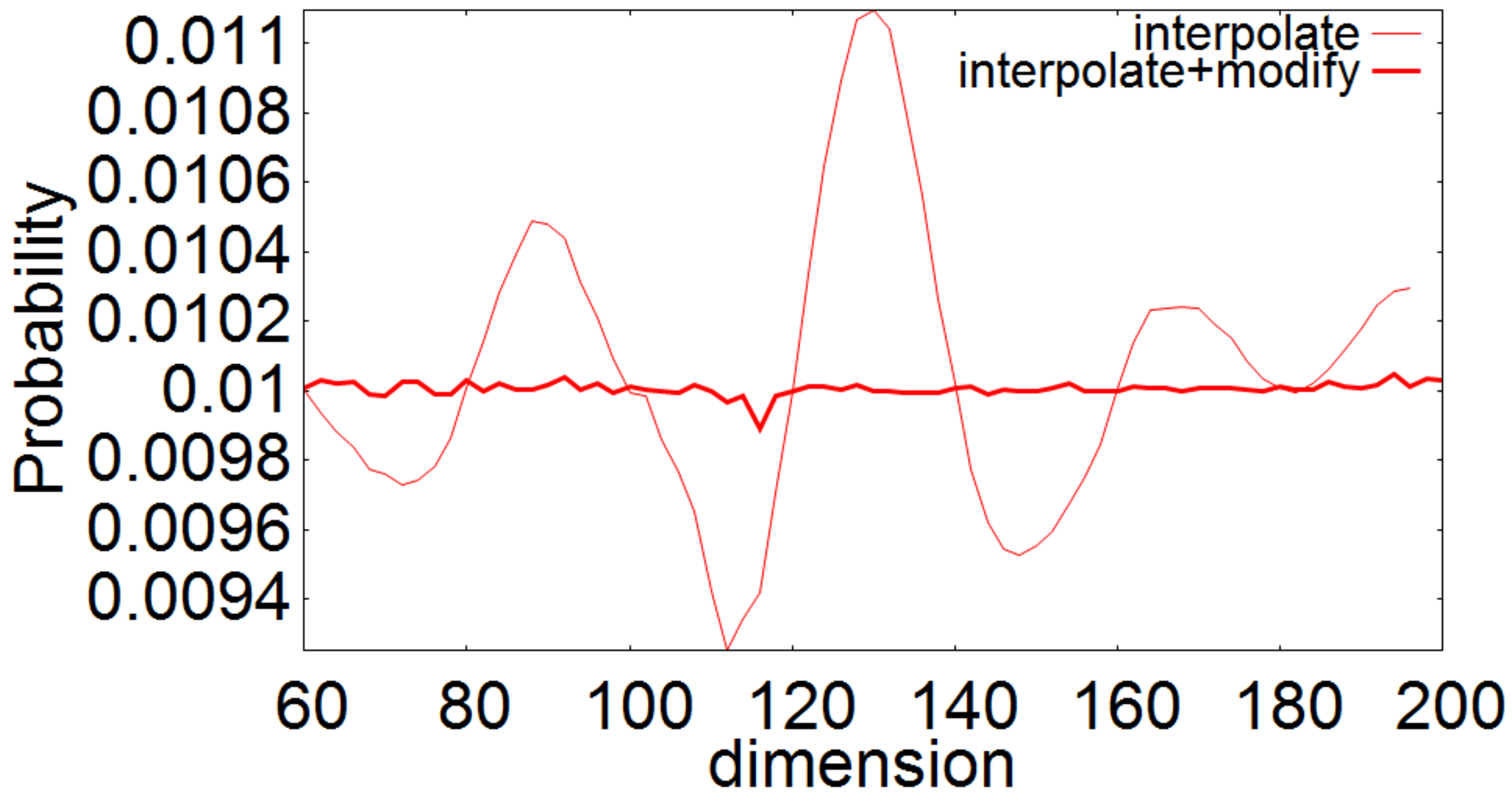}
	\caption{Probabilities of interpolated and interpolate-then-modified functions \label{fig:interpolate probs}}
\end{figure}

From the experimental values, 
it is possible to compute the curves for any parameters by interpolation,
whereas a simple spline interpolation makes curves 
whose probabilities are far from the target (the thin line Fig.~\ref{fig:interpolate probs}).
Modifying them, 
we can obtain the curve whose probability has an error rate of less than 1\%.
Throughout this section, 
we use the word {\it interpolation} in the meaning of spline interpolation.

\medskip
\noindent
{\bf Simple interpolation of the optimized curves}:
Consider the defining points as a four-variable function $s[i,n,\delta,p_0]$.
Since the values of several points are known, 
the other values are computed by the interpolation.
First for each $i,\delta$ and $p_0$ such that values are computed,
interpolate the functions using the values of $n=60,80,\ldots,200$.
Then fix $i,n,p_0$ such that the values are computed, interpolate them as a function of variable $\delta$.
Therefore,
a table of roughly optimized curves for any $n=60,61,\ldots,200$ and $\delta \in [1.005,1.02]$ are computed.
However, as shown in
Fig.~\ref{fig:interpolate probs}
whose thin line shows the approximated probability (\ref{eqn:approxprob})
of curves interpolated from the optimized curves of  $(\delta,p_0)=(1.01,0.01)$,
the probabilities do not match the target probability.

\medskip

\noindent
{\bf Modifying the probabilities of the curve}:
By the method in Section~\ref{sec:mainprobcomp},
for each optimized curve for parameter tuple $(n,\delta,p_0)$,
that is, except for the interpolated parameters, 
precompute the lower and upper bound of the probability.
Let them be $L$ and $U$,
and approximated probability be $\widetilde{p_0}$.
Then, we precompute the {\it modifying constant} $a$ defined as the value satisfying $\widetilde{p_0} = aL + (1-a)U$,
and for other parameters, again interpolate $a$.
Since the procedure for computing $L$ and $U$ for odd $n$ 
differs from that for even $n$,
it needs to precompute modifying constants for several odd $n$.
From the interpolated pruning coefficients for  $n=61,81,101,121,141,161,181$ and $199$,
we precompute the table of $a$.
Table~\ref{tab:modifyingconstants} shows the example values of $a$ for $\delta=1.01$, $p_0=0.01$.

Our method to modify the curve is given as follows.
Let the target probability be $p$.
Suppose two curves for parameters $(n,\delta,p_U)$ and $(n,\delta,p_L)$ such that $10^{-e_L}=p_L \le p < p_U = 10^{-e_U}$ 
are computed (interpolated), and let them be $(R_1^U,\ldots,R_n^U)$ and $(R_1^L,\ldots,R_n^L)$, respectively.
Let the modifying constants for the curve be $a_L$ and $a_U$.
When $p_0>0.1$, we use $p_U=1$, $R_i=1$ for $i\in [n]$, and $a_U=a_L$.
Then merge the curve as $R^2_i = c(R_i^L)^2 + (1-c)(R_i^U)^2$ for $c \in [0,1]$,
and search the suitable $c$ giving desired probability by binary search.
Here, to obtain an approximation of the probability of merged curve, we also set the modifying constant as 
$a=c\cdot a_L + (1-c)\cdot a_U$ and compute the approximated probability as $p=aL+(1-a)U$.
Noting that we also modify the final $R_i$ to a monotonically increasing curve between 0 and 1.
The modification process can be finished by a few numbers of volume computations of 
truncated polygons.
Our implementation needs about 0.08 sec. (60 dim.) to 2.5 sec (200 dim) to output the curve.

\begin{table}
	\caption{Modifying constants for $\delta=1.01$ and  $p_0=0.01$\label{tab:modifyingconstants}}
	\begin{center}
	\begin{footnotesize}
	\begin{tabular}{|l||l|l|l|l|l|}

		\hline
		$n$&60&80&100&120&140 \\
		\hline
		$a$&0.3581&0.3402&0.3342&0.3252&0.3200 \\
		\hline
		\hline
		$n$&61&81&101&121&141 \\
		\hline
		$a$&0.4858&0.4775&0.4668&0.4542&0.4514\\
		\hline
	\end{tabular}
	\end{footnotesize}
	\end{center}
\end{table}

Fig.~\ref{fig:interpolate probs} shows comparison between the interpolated probability and 
modified probability.
While the interpolations have about a 10\% error rate,
the modified probability has an  error rate below 1\%.

\begin{figure}[h]
\includegraphics[scale=0.25]{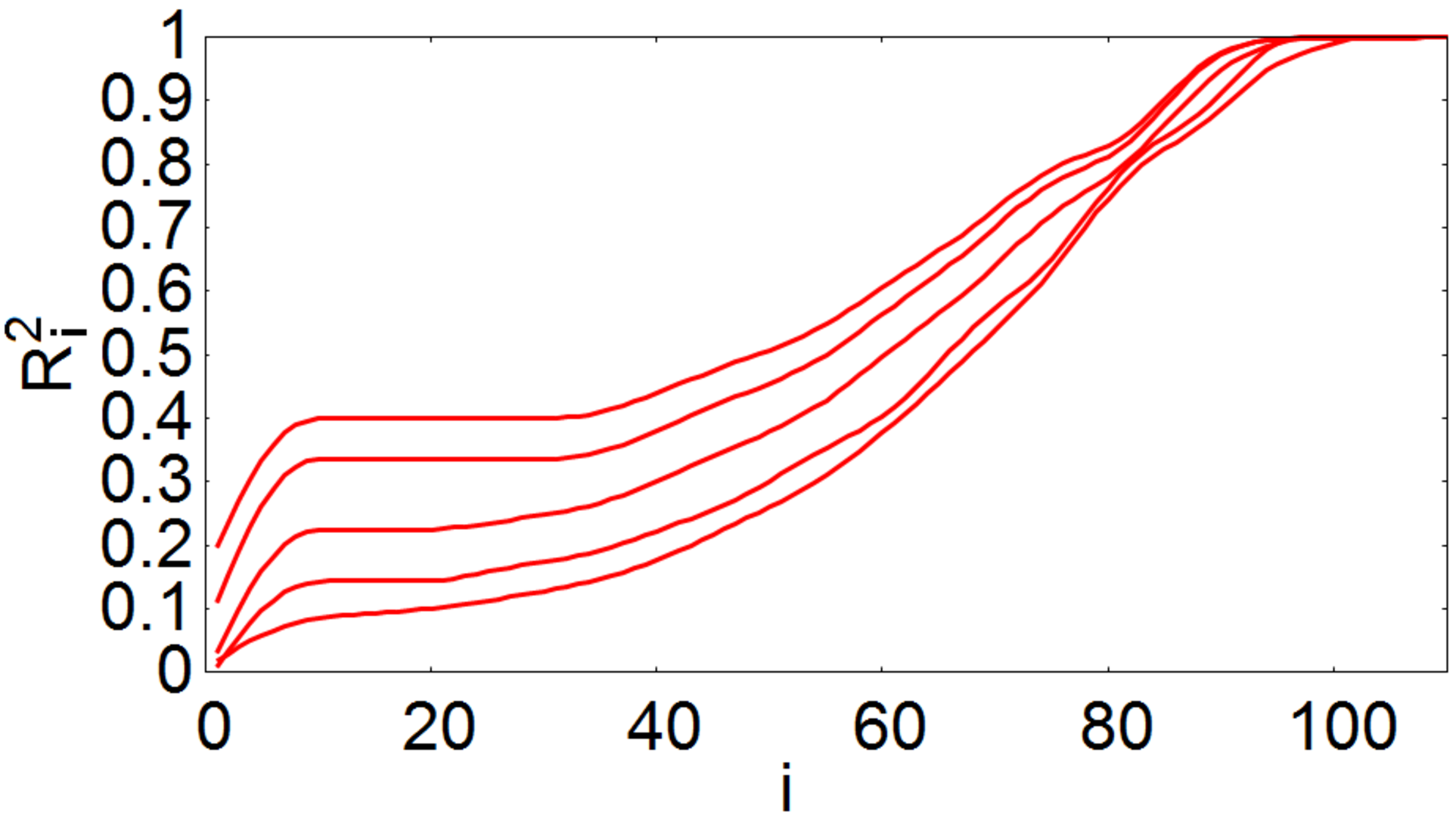}
\caption{Generated curves for $\delta=1.0128$, $n=110$, and several target probabilities \label{fig:generated_curves}}
\end{figure}

Fig.~\ref{fig:generated_curves} shows the interpolated curves
for $\delta=1.0128$ which corresponds to BKZ-20 \cite{GN08b},
$n=110$, and $p_0=0.6\cdot 2^{-f}$ for $f=0,1,4,10$, and $15$.

\subsection{Comparison with Scaled GNR's Coefficients}

\begin{figure}[h]
	\includegraphics[scale=0.20]{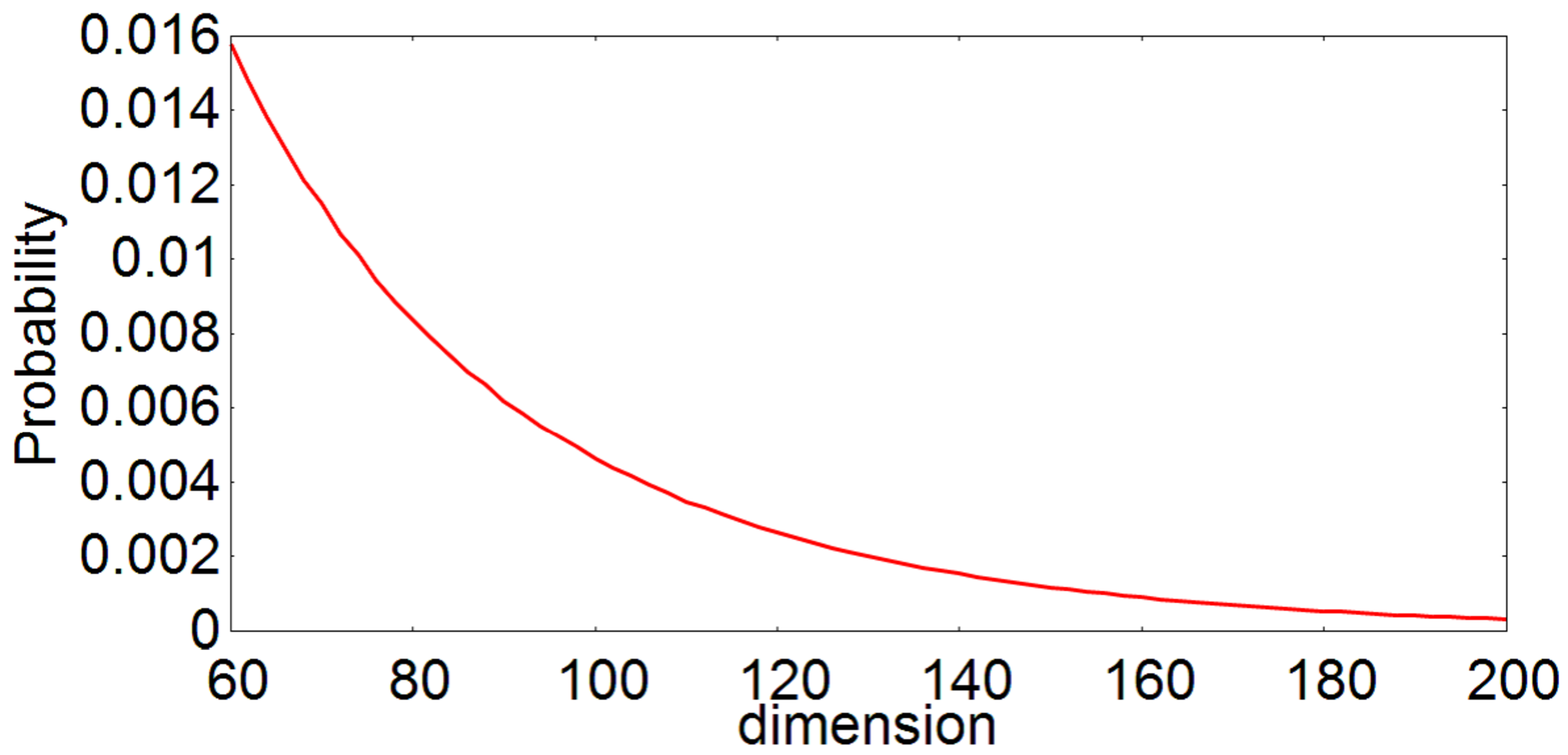}
	\caption{Success probability of scaled GNR \label{fig:success_probs_gnr}}
\end{figure}

We compare our pruning coefficients and 
the scaled version of their optimized function given in \cite[Fig. 1]{GNR10}.
They claimed their function is optimized for a reduced basis of a 110-dimensional knapsack lattice and its success probability is 0.01,
whereas they did not give $||\bold{b}^*_i||$.
In previous works \cite{KSD+11}, they fit the GNR coefficients to a polynomial $f(x)$ defined in $[0,1]$, 
and set their function for $m$-dimensional lattice by  $R^2_i := f(110i/n)$.
Fig.~\ref{fig:success_probs_gnr} shows the success probability of scaled GNR's coefficients.

\begin{figure}[h]
	\includegraphics[scale=0.2]{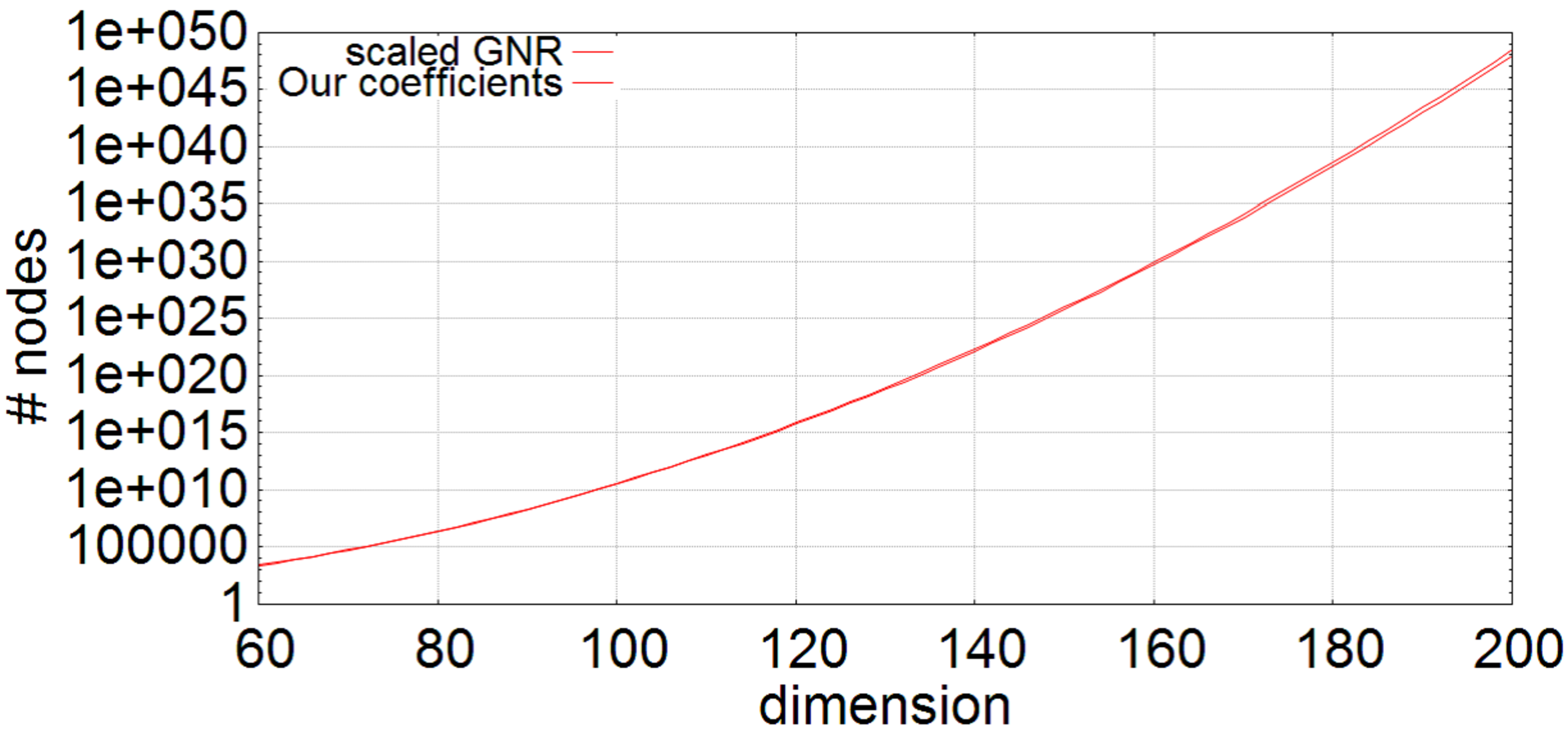}
	\includegraphics[scale=0.25]{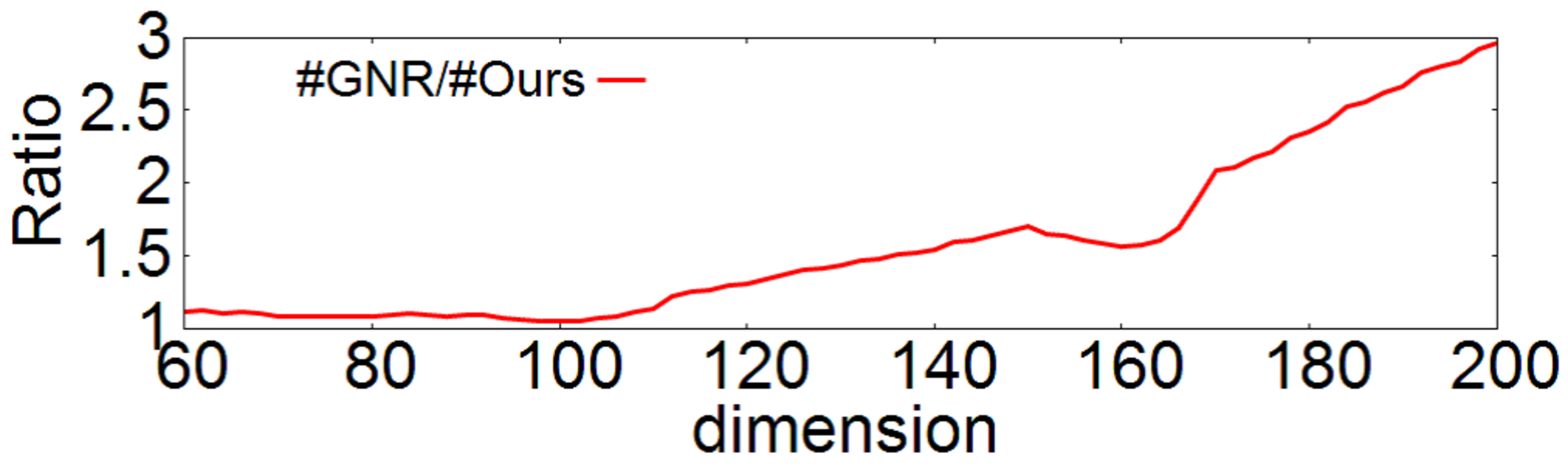}
	\caption{Top: Comparison between number of nodes of scaled GNR and our coefficients; 
	            Bottom: Ratio $\sharp$GNR/$\sharp$Ours \label{fig:comparisonnodes}}
\end{figure}

Fig.~\ref{fig:comparisonnodes} shows a comparison between the number of nodes
for a basis with $\delta=1.0128$.
The pruning coefficient is scaled GNR and ours generated so that the success probability is equal to the others.
Because they are very close to each other, 
we also give the ratio in the bottom graph 
which shows our coefficient is slightly better than others.

Although their coefficients are not optimized for a basis with GSA, 
it would be close to the optimal.
Hence, our interpolated curves are also sufficiently close to the optimized one.

\bigskip

\noindent
{\bf Acknowledgment}:
I would like to thank Phong Q. Nguyen for helpful comments
for writing Section~2.2.
This work was supported by
Grant-in-Aid for Scientific Research on Innovative Areas
Grant Number 24106007.

\bibliographystyle{plain}
\bibliography{refaono}

\end{document}